\documentclass[twocolumn,english,pra, aps, twocolumn, superscriptaddress]{revtex4-1}
\usepackage[T1]{fontenc}
\usepackage[latin9]{inputenc}
\usepackage{color}
\usepackage{mathrsfs}
\usepackage{amstext}
\usepackage{amssymb}
\usepackage{graphicx}
\usepackage{esint}

\makeatletter
\usepackage{xcolor}
\usepackage[colorlinks,citecolor=red,linkcolor=blue]{hyperref}

\makeatother

\usepackage{babel}
\begin{document}
\title{Finite-Time Optimization of Quantum Szilard heat engine}
\author{Tan-Ji Zhou}
\address{Graduate School of China Academy of Engineering Physics, No. 10 Xibeiwang
East Road, Haidian District, Beijing, 100193, China}
\address{Department of Physics, Peking University}
\author{Yu-Han Ma}
\email{yhma@bnu.edu.cn}

\address{Department of Physics, Beijing Normal University, Beijing 100875,
China}
\address{Graduate School of China Academy of Engineering Physics, No. 10 Xibeiwang
East Road, Haidian District, Beijing, 100193, China}
\author{C. P. Sun}
\email{suncp@gscaep.ac.cn}

\address{Graduate School of China Academy of Engineering Physics, No. 10 Xibeiwang
East Road, Haidian District, Beijing, 100193, China}
\address{Beijing Computational Science Research Center, Beijing 100193, China}
\begin{abstract}
We propose a finite-time quantum Szilard engine (QSE) with a quantum
particle with spin as the working substance (WS) to accelerate the
operation of information engines. A Maxwell's demon (MD) is introduced
to probe the spin state within a finite measurement time $t_{{\rm M}}$
to capture the which-way information of the particle, quantified by
the mutual information $I(t_{\rm{M}})$ between WS and MD. We establish
that the efficiency $\eta$ of QSE is bounded by $\eta\leq1-(1-\eta_{\rm{C}}){\rm ln}2/I(t_{{\rm M}})$,
where $I(t_{{\rm M}})/\rm {ln}2$ characterizes the
ideality of quantum measurement, and approaches $1$ for the Carnot
efficiency reached under ideal measurement in quasi-static regime.
We find that the power of QSE scales as $P\propto t_{{\rm M}}^{3}$ in the
short-time regime and as $P\propto t_{\rm M}^{-1}$ in the long-time regime.
Additionally, considering the energy cost for erasing the MD's memory
required by Landauer's principle, there exists a threshold time that
guarantees QSE to output positive work.
\end{abstract}
\maketitle
\textit{\textcolor{blue}{Introduction.}}---The field of information
heat engines \citep{1929Szilard,2009MaruyamaRMP.81.1} was developed
to establish the fundamental role of information in thermodynamics
\citep{2009MaruyamaRMP.81.1,2015thermodynamicsinformation}. These
engines operate by generating work through information processing,
rather than conventional heat absorption. A well-known example is
the Szilard engine \citep{1929Szilard}, which involves the Maxwell's
demon (MD) probing the position of particles in a single heat reservoir
to extract work. The apparent violation of the second law of thermodynamics
in this case is resolved \citep{quan2006maxwell,2009MaruyamaRMP.81.1,dong2011quantum,kim2011quantum,cai2012multiparticle,2015thermodynamicsinformation}
by taking into account the energy cost of erasing the demon's memory,
as per Landauer's principle \citep{1961Landauer5392446}. Several
quantum Szilard engines (QSEs) have been proposed \citep{quan2006maxwell,2009MaruyamaRMP.81.1,dong2011quantum,kim2011quantum,cai2012multiparticle,2015thermodynamicsinformation}
to explore the connection between information and thermodynamics in
quantum regime. 
\begin{figure}[h]
\centering{}\includegraphics[width=8.5cm]{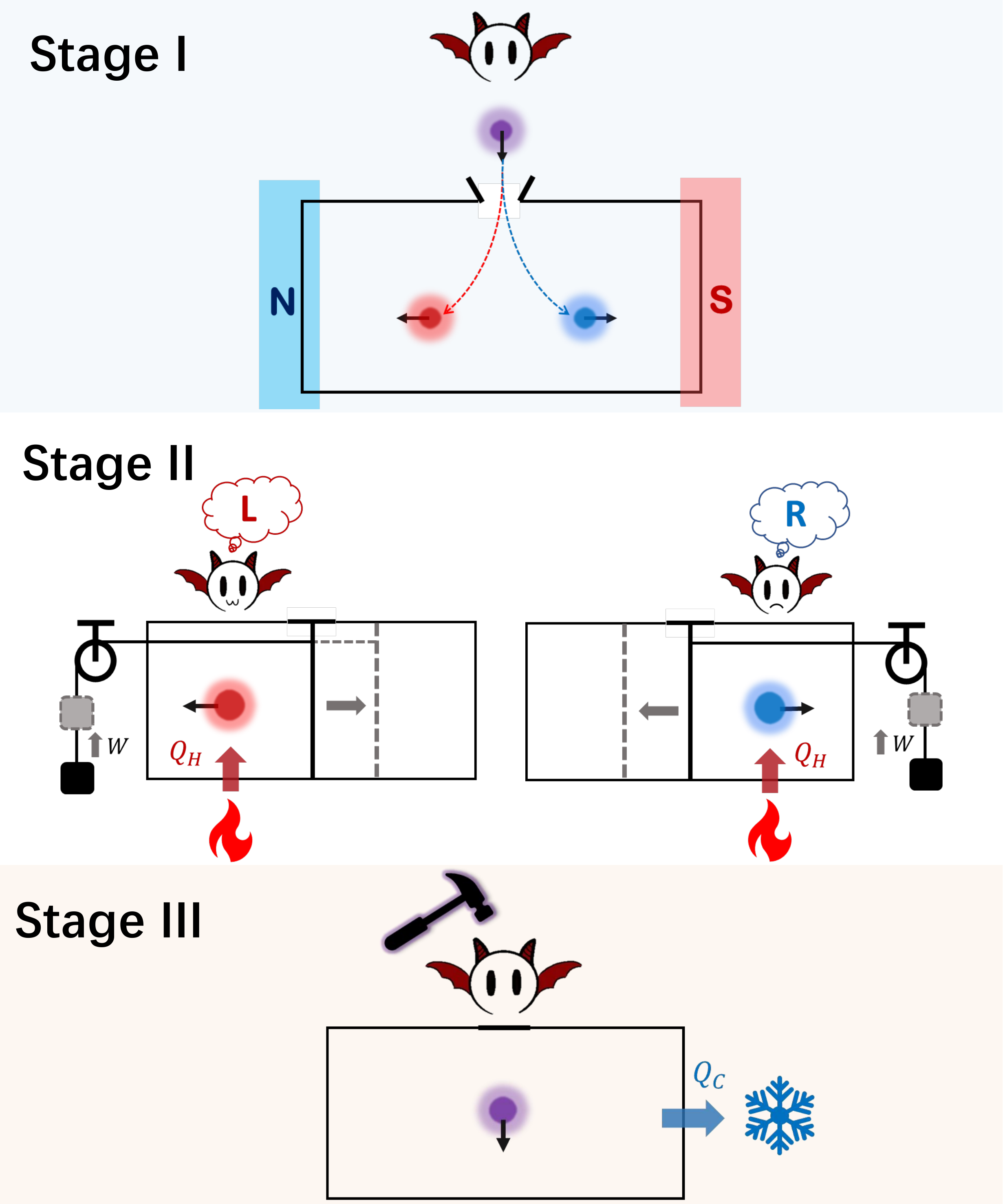}\caption{\label{fig:Schematic-diagram-of}Schematic illustration of the Szilard
engine cycle. MD first observes the position of the particle (through
the detection of the particle's spin) in the box (stage I). Then proper
operation is chosen to extract work from collisions of particle and
baffle (stage II). After the work done process, the information of
the particle's position recorded by MD is erased, and the entire system
returns to its initial state (stage III).}
\end{figure}

In QSEs, the demon's recording of the particle's position is equivalent
to the establishment of entanglement between MD and the particle.
However, high-speed cyclic operation needs to be accomplished within
a short measurement time to achieve optimal power of QSE. This makes
it impossible to carry out an ideal quantum measurement, according
to the decoherence theory for quantum measurement \citep{joos2013decoherence,zeh1970interpretation,zurekDecoherenceEinselectionQuantum2003,schlosshauerDecoherenceMeasurementProblem2005}.

In this Letter, we propose and study a finite-time QSE model to analyze
the influence of non-ideal quantum measurement on engine performance.
We investigate the information correlation between the particle and
MD during quantum measurement and analyze the dynamics of the which-way
information recorded by MD. Our results demonstrate the fundamental
trade-off between the power and efficiency of QSE due to non-ideal
measurement. Although information recording does not consume additional
energy, the non-ideality of measurement affects the thermodynamic
signatures of the QSE. 

\textit{\textcolor{blue}{Operation cycle of a QSE.}}---As illustrated
in Fig. \ref{fig:Schematic-diagram-of}, the working substance (WS)
of the considered QSE is specified as a single particle with spin-1/2.
The MD probes the particle's position through the detection of its
spin state to obtain the which-way information of the particle, which
can be further used to output work. The operation cycle of the QSE
involves three stages, namely, information recording through quantum
measurement (stage I), work outputting (stage II) and information
erasing (stage III).

\textbf{In stage I}, the particle is injected into a cylinder, and
the inhomogeneous magnetic field which perpendicular to the direction
of the particle's incidence is turned on. This results in the entanglement
of the particle's spin and spatial states, enabling the Maxwell's
demon to detect the particle's position through measurement of its
spin state.

After stage I, the spin-magnetic field interaction gives rise to a
time-dependent correlation $I(t_{M})$ between the spin state and
the spatial state, and the deflection of the particle can be inferred
from the spin detection. Hereafter, the duration of stage I $t_{M}$
is called the measurement time. We qualify such correlation with the
mutual information \citep{kullback1997information} as

\begin{equation}
I(t_{\rm{M}})=I_{\rm{DP}}\equiv S_{{\rm D}}+S_{{\rm p}}-S_{{\rm DP}},
\end{equation}
where $S_{{\rm DP}}$, $S_{{\rm D}}$ and $S_{{\rm p}}$ are the Shannon
entropy of the total system, the MD and the particle, respectively.

\textbf{In stage II}, the magnetic field is turned off and a suitable
protocol is chosen according to MD's memory to enable the extraction
of work from collisions of the particle and baffle. The engine is
contacting with a heat reservoir with temperature $T_{{\rm H}}$ when
it outputs work. At the end of stage II, the correlation between the
particle and MD vanished due to their thermalization in the reservoir,
thereby contributing to an increase of their total entropy, i.e.,
$\Delta S_{{\rm DP}}=I(t_{\rm{M}})$ \citep{li2017production,Still2020PRL.124.050601}.

The generalized free energy \citep{crooks2007beyond,Sagawa2008PRLFreeEnergy,takara2010generalization,lostaglio2015description,deffner2016quantum,Still2020PRL.124.050601}
of the total system containing the particle and MD

\begin{equation}
F\equiv U-k_{{\rm B}}T_{{\rm H}}S_{{\rm DP}}\label{eq:F}
\end{equation}
is defined by Shannon entropy $S_{{\rm DP}}$ instead of the thermodynamic
entropy in conventional cases. Here, $U$ is the energy of the total
system. Consequently, it follows from Eq. (\ref{eq:F}) and the 2nd
law of thermodynamics ($W\text{\ensuremath{\leq-\Delta F}}$) that
the output work is bounded above by the spin-position correlation
of the particle as

\begin{equation}
W_{{\rm O}}\leq k_{{\rm B}}T_{{\rm H}}I(t_{M}),\label{eq:Wo}
\end{equation}
where the facts that the internal energy of an ideal gas particle
is unchanged in the isothermal process and MD's energy remains unchanged
throughout the whole cycle have been used. As a result, heat in the
amount of $Q_{{\rm O}}=W_{{\rm O}}$ is absorbed from the heat reservoir.

\textbf{In stage III}, the information recorded by MD is erased, and
the entire system returns to its initial state. In this process, the
engine is in contact with another reservoir with temperature $T_{{\rm C}}$.
According to Landauer\textquoteright s principle, work in the amount
of $W_{{\rm E}}\geq k_{{\rm B}}T_{{\rm C}}{\rm ln}2$ must beapplied
to the system \citep{1961Landauer5392446} to erase MD's memory (to
initialize the spin state in our case). The same amount of heat $Q_{{\rm E}}=W_{{\rm E}}$
is dissipated into the reservoir according to the first law of thermodynamics
in this isothermal process.

\textit{\textcolor{blue}{Realization of the cycle.}}---Taken Stern-Gerlach
experiment \citep{gerlach1922experimentelle,bohm2012quantum,platt1992modern,griffithsIntroductionQuantumMechanics2018}
as an realization of stage I, the correlation between spin state and
the spatial state of the particle can be specifically obtained (See
Sec. I of the Supplementary Materials (SM) for details \citep{S-M}).
We define the measurement ideality as $\text{\ensuremath{\mathcal{M}\equiv I/{\rm ln}2}}$
and it is explicitly written as \citep{S-M}

\begin{equation}
\mathcal{M}(\tilde{t})=1+\frac{p(\tilde{t}){\rm ln}p(\tilde{t})+\left[1-p(\tilde{t})\right]{\rm ln}\left[1-p(\tilde{t})\right]}{{\rm ln}2}
\end{equation}
with

\begin{equation}
p(\tilde{t})=\frac{1}{2}\left[1+{\rm erf}\left(\frac{\alpha\tilde{t}^{2}}{\sqrt{2\tilde{t}^{2}+8}}\right)\right]
\end{equation}
the possibility that MD can accurately infer the particle's position.
Here, ${\rm erf}(z)=2\pi^{-1/2}\int_{0}^{z}\mathrm{exp}(-u^{2})\text{{\rm d}}u$
is the Gaussian error function, $\tilde{t}\equiv t_{\rm{M}}/\tau_{\rm{M}}$
is the dimensionless measurement time. The characteristic time of
the measurement process $\tau_{\rm{M}}\equiv ma^{2}/\hbar$ is determined
by the particle mass $m$ and the width of the particle's spatial
wave packet $a$. In addition, $\text{\ensuremath{\alpha}}\equiv ma^{3}f/\hbar^{2}$
is proportional to the magnetic force $f$ acting on the spin.

\begin{figure}
\begin{raggedright}
\includegraphics[width=4cm]{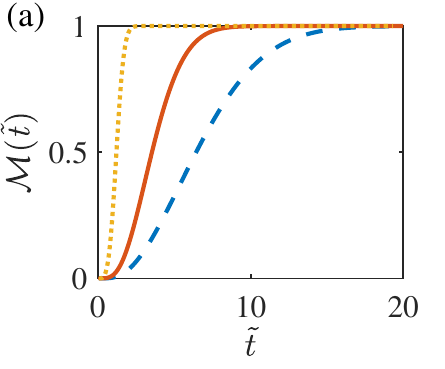}\includegraphics[width=4cm]{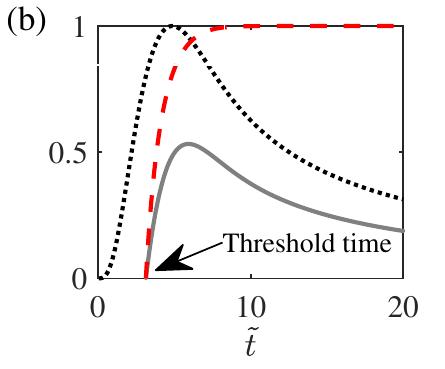}
\par\end{raggedright}
\begin{centering}
\includegraphics[width=8.5cm]{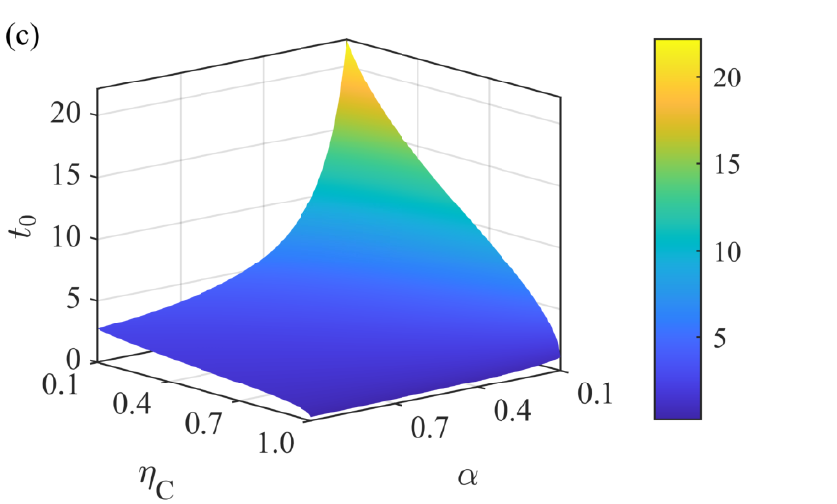}
\par\end{centering}
\centering{}\caption{\label{fig:Normalized-Ideality_Efficiency-and-power}(a) Measurement
ideality$\text{\ensuremath{\mathcal{M}(t)=I(t)/{\rm ln}2}}$ with
different $\alpha$. The blue dashed curve, orange solid curve, and
yellow dotted curve are plotted with $\alpha=0.2$, $\alpha=0.4$,
and $\alpha=2$, respectively. (b) Efficiency and power of the engine
as the function of dimensionalized measurement time, where $\eta_{{\rm C}}=0.6$,
$\alpha=0.4$ are used. The grey solid curve and the black dotted
curve mark $P(\tilde{t})$ and $P_{\mathrm{O}}(\tilde{t})$ respectively,
which are normalized by $\max\left[P_{\mathrm{O}}(\tilde{t})\right]$,
and the red dashed line marks $\eta(\tilde{t})$ which is normalized
by $\eta_{\mathrm{C}}$. (c) The threshold time $t_{0}$ as a function
of the Carnot efficiency $\eta_{\mathrm{C}}$ and the characteristic
parameter $\text{\ensuremath{\alpha}}=ma^{3}f/\hbar^{2}$. The yellow
area corresponds to higher value of $t_{0}$ and the blue area corresponds
to the lower part. $t_{0}$ is numerically solved from Eq. (\ref{eq:W>0})
for each given set of $\eta_{{\rm C}}$ and $\text{\ensuremath{\alpha}}$.}
\end{figure}

We further plot $\mathcal{M}(\tilde{t})$ as a function of $\tilde{t}$
in Fig. \ref{fig:Normalized-Ideality_Efficiency-and-power} (a). It
is clearly seen in this figure that for a given measurement time,
larger $\alpha$ results in a greater degree of correlation generated
between the spin degrees of freedom and the spatial degrees of freedom
of the particle, and thus MD can more accurately distinguish the position
of the particle. As the measurement time increases, the spin-position
correlation of the particle monotonically increases. In the ideal
measurement process, we have $\lim_{\tilde{t}\rightarrow\infty}\mathcal{M}(\tilde{t})=1$
($p(\tilde{t})\rightarrow1$), which is the plateau approached by
the three curves in Fig. \ref{fig:Normalized-Ideality_Efficiency-and-power}(a).

We emphasize here that the essential difference between the Szilard
engine with finite-time quantum measurement proposed in this work
and the conventional one is reflected in the maximum extracted work
allowed by the information recording process. In the conventional
Szilard engine, MD can deterministically observe the position of the
particle, enabling a precise judgment on which operation to choose
to extract work. However, in our case, the dynamics of the measurement
renders the inference of the particle's position from its spin state
probabilistic. The possibility of selecting an incorrect operation
for extracting work will reduce the output work, which has been recently
discussed for the correlated variable Szilard engine in non-dynamical
case \citep{Still2020PRL.124.050601}.

In particular, $\tilde{t}\rightarrow0$ is associated with invalid
measurement with $\mathcal{M}\rightarrow0$. Consequently, MD knows
nothing about the particle's position, and thus no work can be extracted.
On the other hand, when an ideal measurement with $\mathcal{M}=1$
is achieved in the quasi-static regime ($\tilde{t}\rightarrow\infty$),
Eq. (\ref{eq:Wo}) recovers the result $W_{{\rm O}}^{{\rm ideal}}=k_{{\rm B}}T_{{\rm H}}\ln2$
of the conventional Szilard engine. Hence we can rewrite the upper
bound of the outputting work in Eq. (\ref{eq:Wo}) as $W_{{\rm O}}\leq\mathcal{M}W_{{\rm O}}^{{\rm ideal}}.$

\textit{\textcolor{blue}{Efficiency and power of the QSE.}}---With
the energy conversion relations in the above cycle analysis in mind,
the total work of the QSE performing in one cycle reads

\begin{equation}
W=W_{{\rm O}}-W_{{\rm E}}\leq\mathcal{M}W_{{\rm O}}^{{\rm ideal}}-k_{{\rm B}}T_{{\rm C}}{\rm ln}2,\label{eq:work}
\end{equation}
and thus the efficiency of the QSE $\eta\equiv W/Q_{{\rm O}}=W/W_{{\rm O}}$
satisfies

\begin{equation}
\eta\leq1-\frac{k_{{\rm B}}T_{{\rm C}}{\rm ln}2}{\mathcal{M}W_{{\rm O}}^{{\rm ideal}}}=1-\frac{\left(1-\eta_{{\rm C}}\right).}{\mathscr{\mathcal{M}}}\label{eq:eta}
\end{equation}
The above inequality can be reorganized in terms of the Carnot efficiency
$\eta_{{\rm C}}\equiv1-T_{{\rm C}}/T_{{\rm H}}$ as

\begin{equation}
\frac{1-\eta_{{\rm C}}}{1-\eta}\leq\mathcal{M},\label{eq:eta_ineq}
\end{equation}
which is the main result of this Letter, revealing fundamental upper
bound on energy conversion efficiency set by information measurement
ideality. Here, the equal sign is saturated if and only if there is
no irreversibility in stage II and stage III.

By definition, the power of the QSE is $P\equiv W/t_{{\rm cycle}}$,
where the cycle time is $t_{{\rm cycle}}=t_{{\rm M}}+t_{{\rm O}}+t_{{\rm E}}$
with the duration of the work outputting (information erasing) process
to be $t_{{\rm O}}$ ($t_{{\rm E}}$). In order to significantly demonstrate
the impact of finite-time measurements on the performance of QSE,
we first consider that the coupling between the system and measurement
instrument is weak enough so that the finite-time behavior of the
information recording dominates in the considered time scale ($t_{{\rm E}},t_{{\rm O}}\ll t_{{\rm M}}$)
\citep{S-M}. In this case, the irreversibility of the work outputting
and information erasing processes can be ignored, and the power of
the engine reduces to

\begin{equation}
P=\frac{W}{t_{{\rm M}}}\leq\frac{\mathcal{M}W_{{\rm O}}^{{\rm ideal}}-k_{{\rm B}}T_{{\rm C}}{\rm ln}2}{t_{{\rm M}}}=\frac{\eta(\tilde{t})k_{{\rm B}}T_{{\rm H}}I(\tilde{t})}{\tau_{{\rm M}}\tilde{t}}.
\end{equation}
We further define the output power associated with the work outputting
process as $P_{{\rm O}}(\tilde{t})\equiv W_{{\rm O}}/t_{{\rm M}}=k_{{\rm B}}T_{{\rm H}}I(\tilde{t})/(\tau_{{\rm M}}\tilde{t})$,
where the equal sign in Eq. (\ref{eq:eta_ineq}) is taken.

We illustrate $\eta(\tilde{t})$, $P(\tilde{t})$, and $P_{{\rm O}}(\tilde{t})$
as functions of non-dimensionalized measurement time $\tilde{t}$
in Fig. \ref{fig:Normalized-Ideality_Efficiency-and-power} (b), where
$P(\tilde{t})$ and $P_{{\rm O}}(\tilde{t})$ are non-dimensionalized
by the maximum value of $P_{{\rm O}}(\tilde{t})$, and $\eta(\tilde{t})$
is normalized by $\eta_{\mathrm{C}}$. As illustrated in this figure,
$P_{{\rm O}}(\tilde{t})$ increases from zero with the increase of
measurement time $\tilde{t}$ until it reaches its maximum value and
then decreases gradually, which is positive for all $\tilde{t}>0$.
In the short time regime, $\mathcal{M}(\tilde{t})\propto\alpha^{2}\tilde{t}^{4}$
\citep{S-M}, and thus $P_{{\rm O}}(\tilde{t}\ll1)\propto\alpha^{2}\tilde{t}^{3}$,
which indicates the output power is proportional to the square of
the strength of measurement ($\alpha^{2}$). On the other hand, in
the long time regime, $\mathcal{M}(\tilde{t})$ tends to $1$ exponentially
(see Fig. \ref{fig:Normalized-Ideality_Efficiency-and-power} (a))
such that $P_{{\rm O}}(\tilde{t}\gg1)\propto\tilde{t}^{-1}$.

\textit{\textcolor{blue}{Threshold time for positive work.}}--- It
is observed in Fig. \ref{fig:Normalized-Ideality_Efficiency-and-power}
(b) that the power $P(\tilde{t})$ and the efficiency $\eta(\tilde{t})$
of the information heat engine are positive only when the time exceeds
a threshold value. This can be understood from Eq. (\ref{eq:work})
that the positive work condition $W\geq0$ sets a lower bound for
measurement ideality as

\begin{equation}
\mathcal{M}(\tilde{t})\geq1-\eta_{{\rm C}}.\label{eq:W>0}
\end{equation}
Naturally, we define the threshold time $t_{0}\equiv\mathcal{M}^{-1}(1-\eta_{{\rm C}})$
as the lower bound of the dimensionless measurement time that holds
the above inequality. The positive work condition for the engine is
thus converted to a restriction on the measurement time $\tilde{t}>t_{0}$.

For different value of $\eta_{{\rm C}}$, the shape of the curves
in Fig. \ref{fig:Normalized-Ideality_Efficiency-and-power} (b) will
change on the premise that the overall characteristics remain unchanged,
resulting in different values of $t_{0}$. We numerically solve $t_{0}$
from Eq. (\ref{eq:W>0}) for each given set of parameters $\left\{ \eta_{{\rm C}}\in(0.1,1),\ensuremath{\alpha\in(0.1,1)}\right\} $,
and then plot $t_{0}$ as a function of $\eta_{\mathrm{C}}$ and $\text{\ensuremath{\alpha}}$
in Fig. \ref{fig:Normalized-Ideality_Efficiency-and-power} (c). This
figure shows that the threshold time increase remarkably as $\alpha$
grows beyond a certain value, while it decreases rather gradually
with the increase of $\eta_{{\rm C}}$ for relatively large $\text{\ensuremath{\alpha}}$.
Therefore, we can increase $\alpha$ to achieve a stronger measurement,
thereby effectively reducing $t_{0}$. By doing this, the engine can
output positive work in a shorter cycle time. In other words, increasing
$\alpha$ can speed up the engine operation without decreasing the
output work, thereby improving the power of the engine.

\begin{figure}
\begin{raggedright}
\includegraphics[width=8.5cm]{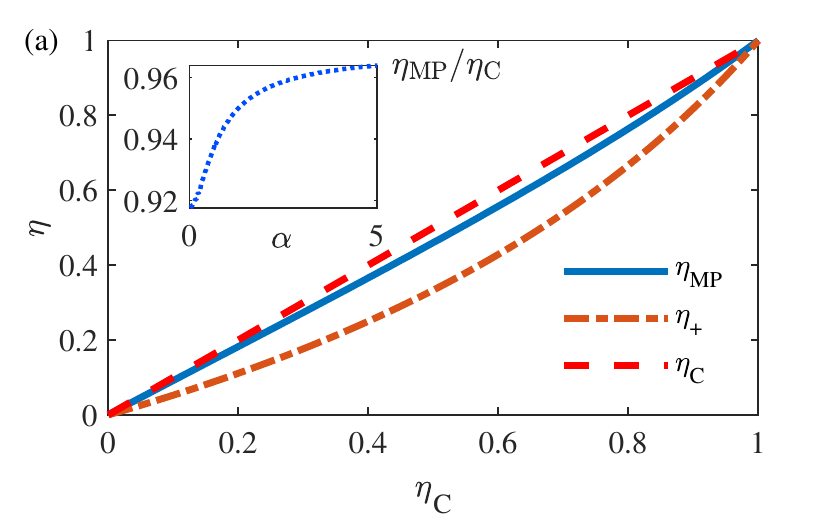}
\par\end{raggedright}
\begin{raggedright}
\includegraphics[width=8.5cm]{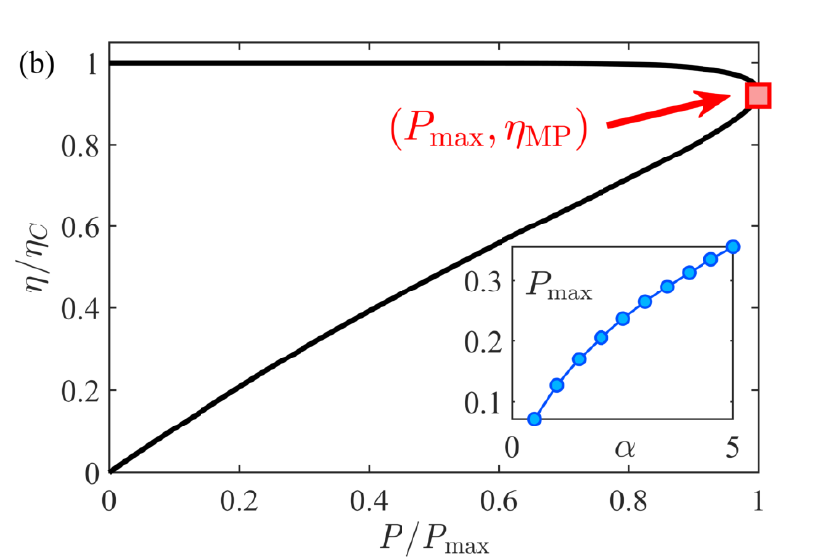}
\par\end{raggedright}
\centering{}\caption{\label{fig:EMP} (a) Efficiency at maximum power of QSE as a function
of $\eta_{C}$ . The blue solid curve represents the efficiency at
maximum power (EMP) of the information engine, the typical bound of
EMP $\eta_{+}=\eta_{\rm{C}}/(2-\eta_{\rm{C}})$ is plotted
with the orange dash-dotted curve, and the red dashed line marks the
Carnot efficiency $\eta_{\rm{C}}=1-T_{\rm{C}}/T_{\rm{H}}$,
in this plot $\alpha=0.4$ is used. In the inset figure, EMP as a
function of $\alpha$ is plotted with the blue dotted curve, and $\eta_{{\rm C}}=0.6$
is used. (b) Power-efficiency trade-off relation of QSE with $\alpha=0.4$
and $\eta_{{\rm C}}=0.6$. In this figure, only the finite-time effect
of the information recording process (stage I) is considered. In the
inset figure, the maximum power $P_{\rm{max}}$ as a function
of $\alpha$ is plotted with the blue circle-dotted curve.}
\end{figure}

\textit{\textcolor{blue}{Optimal operation of the QSE.}}--- When
the power of the information heat engine $P(\tilde{t})$ reaches its
maximum at the measurement time $\tilde{t}=\tilde{t}^{*}$, i.e.,
$P(\tilde{t}^{*})={\rm max}\left\{ P(\tilde{t})|\tilde{t}>0\right\} \equiv P_{{\rm max}}$,
the corresponding efficiency is referred to as the efficiency at maximum
power (EMP). In Fig. \ref{fig:EMP} (a), the EMP of the engine $\text{\ensuremath{\eta_{{\rm MP}}}}$as
a function of $\eta_{{\rm C}}$ is plotted with the blue solid curve.
The typical efficiency $\eta_{+}=\eta_{\mathrm{C}}/(2-\eta_{\mathrm{C}})$,
known as the upper bound of the EMP of low-dissipation Carnot heat
engine \citep{EspositoPRL2010}, is marked with the orange dash-dotted
curve. It can be observed from Fig. \ref{fig:EMP} that the EMP of
QSE can exceed $\eta_{+}$ and finally approaches the Carnot efficiency
$\eta_{\mathrm{C}}=1-T_{\mathrm{C}}/T_{\mathrm{H}}$ (red dashed line)
as $\eta_{{\rm C}}\to1$. In this example, $\alpha=0.4$ is used.
Note that $\eta_{+}$ is the overall upper bound for EMP of all near-equilibrium
engines \citep{ZCTuCPB}, the significantly higher EMP of our QSE
indicates the quantum thermodynamical advantage away from equilibrium
regime. In addition, the dependence of $\eta_{{\rm MP}}/\eta_{{\rm C}}$
on $\alpha$ is shown in the inset figure, indicating that the EMP
of QSE gradually increases with $\alpha$ and approaches the Carnot
efficiency very closely. 

Moreover, the power-efficiency trade-off relation \citep{holubecMaximumEfficiencyLowdissipation2016,TradeoffrelationShiraishi,CavinaPRLtradeoffrelation,Constraintrelationyhma}
of our QSE is illustrated in Fig. \ref{fig:EMP} (b) with $\alpha=0.4$
and $\eta_{{\rm C}}=0.6$. To obtain this trade-off, $P(\tilde{t})$
and $\eta(\tilde{t})$ are calculated numerically by varying the measurement
time $\tilde{t}\in[1,10^{5}]$. Increasing $\tilde{t}$ results in
an increase in $\eta$, while $P$ first increases to reach its maximum
value and then decreases, as shown in Fig. \ref{fig:Normalized-Ideality_Efficiency-and-power}
(a). The EMP of the engine is associated with the rightmost edge of
the curve (red squared dot), which is pointed by the red arrow. The
maximum power $P_{\mathrm{max}}$ as a function of $\alpha$ is demonstrated
in the inset figure, which indicates that an increase in $\alpha$
will lead to a significant increase in $P_{\mathrm{max}}$.

Overall, the above obtained results demonstrate that the dynamics
of information recording realized by quantum measurement has a significant
impact on the operation of the Szilard engine. Specifically, more
complete information recording (larger $\alpha$) leads to higher
maximum power as well as higher EMP of the engine.

\textit{\textcolor{blue}{Remarks on finite-time effects in stages
II and III.}}\textit{---}In the above discussion, only the finite-time
behavior of the information recording process is taken into account.
When we consider the finite-time effects in both the information recording
and information erasing processes, we find that \textbf{i)} the EMP
of the engine can also exceed the upper limit of the efficiency of
low-dissipation Carnot heat engine at a certain range of $\eta_{{\rm C}}$;
\textbf{ii)} the QSE can operate anywhere within the envelop of the
power-efficiency trade-off, which is quite different from the single-variable
case shown in Fig. \ref{fig:EMP} (b) where the system can only operate
with power and efficiency located on the curve. On the other hand,
if the information recording is complete with an ideal measurement
with $I(t_{{\rm M}})=\ln2$, and the operation time of the measurement
process can be ignored in comparison with that of the work outputting
and information erasing process, then the information heat engine
can be exactly mapped into a Carnot-like heat engine, whose EMP satisfies
$\eta_{\mathrm{C}}/2\leq\eta_{\rm{MP}}\leq\eta_{\mathrm{C}}/(2-\eta_{\mathrm{C}})$.
The detailed calculations and discussion on these issues are given
in Sec. II of the SM \citep{S-M}.

\textit{\textcolor{blue}{Conclusions and outlooks.}}---In summary,
we propose a finite-time quantum Szilard heat engine model. The non-ideal
measurement of the working substance carried out by the demon allows
the engine cycle to run fast with high power at the expense of efficiency.
It is found that the engine is able to output positive work only after
the measurement time reaches a threshold, which reflects the dynamic
competition between the output work achieved by information recording
and the work consumed for erasing information. The optimal performance
of such information engine, characterized with the EMP and the power-efficiency
trade-off relation, is obtained quantitatively. These results demonstrate
that quantum measurements have potentially observable thermodynamic
effects, which have a critical impact on thermodynamic cycles of exchanging
information for energy.

This study bridges the gap between the dynamics of quantum measurement
and non-equilibrium thermodynamics, and shall bring new insights in
providing thermodynamic evidence for identifying different interpretations
of quantum measurement \citep{everett1957relative,griffiths1984consistent,zurekDecoherenceEinselectionQuantum2003,mittelstaedt2004interpretation,allahverdyan2013understanding}.
As potential extensions, the optimal control protocol \citep{yhmaoptimalcontrol,2020Optimal}
and the geometric optimization \citep{Crooks2007,maMinimalEnergyCost2022}
of the cycle, the finite-sized effect of the reservoir \citep{ondrechen1981maximum,izumida2014work,2020-finite-size,yuan2022optimizing}
involved in the work output processes and erase process, the many-body
effect of the working substance \citep{YHMaQPTHE} are expected to
be taken into future consideration. In addition, whether there is
a thermodynamic criterion for objectivity in quantum measurement \citep{zurekDecoherenceEinselectionQuantum2003,zurek2009quantum,riedel2016objective,li2018objectivity}
is a fundamental question worth exploring. Finally, inspired by the
obtained results on thermodynamic signatures of information recording
in quantum measurement, we leave a fascinating and challenging question
to end this paper: \textit{What thermodynamic consequences could be
caused by the finite-time memory of intelligent life.}

\textit{\textcolor{blue}{Acknowledgment}}\textcolor{blue}{.}---This
work is supported by the National Natural Science Foundation of China
(Grants No. 12088101, No. U1930402, and No. U1930403). Y.-H. Ma acknowledges
support from the China Postdoctoral Science Foundation (Grant No.
BX2021030).

\bibliographystyle{apsrev}
\bibliography{FTOSHE}

\end{document}


\title{\textit{\Large{}Supplementary Materials for}{\Large{} ``Finite-Time
Optimization of Quantum Szilard heat engine''}}
\author{Tan-Ji Zhou}
\address{Graduate School of China Academy of Engineering Physics, No. 10 Xibeiwang
East Road, Haidian District, Beijing, 100193, China}
\address{Department of Physics, Peking University}
\author{Yu-Han Ma}
\email{yhma@bnu.edu.cn}

\address{Department of Physics, Beijing Normal University, Beijing 100875,
China}
\address{Graduate School of China Academy of Engineering Physics, No. 10 Xibeiwang
East Road, Haidian District, Beijing, 100193, China}
\author{C. P. Sun}
\email{suncp@gscaep.ac.cn}

\address{Graduate School of China Academy of Engineering Physics, No. 10 Xibeiwang
East Road, Haidian District, Beijing, 100193, China}
\address{Beijing Computational Science Research Center, Beijing 100193, China}

\maketitle
This document is devoted to providing the detailed derivations and
the supporting discussions to the main content of the Letter. In Sec.\ref{sec:Information-Recording},
we discuss the information recording in the Stern-Gerlach experiment
and obtain the spatial wave function, overlapping integral of the
wave functions, the conditional spatial distribution probability and
the correlation (mutual entropy $I(\tilde{t})$) between MD and the
spatial state of the particle. In Sec. \ref{sec:Thermodynamic-signature-of-full-finite-time-description},
we calculate the EMP of the Szilard heat engine in more general cases
with full finite time description of the cycle.

\section{\label{sec:Information-Recording}Information recording in the Stern-Gerlach
experiment}

\begin{figure}
\includegraphics[width=8.5cm]{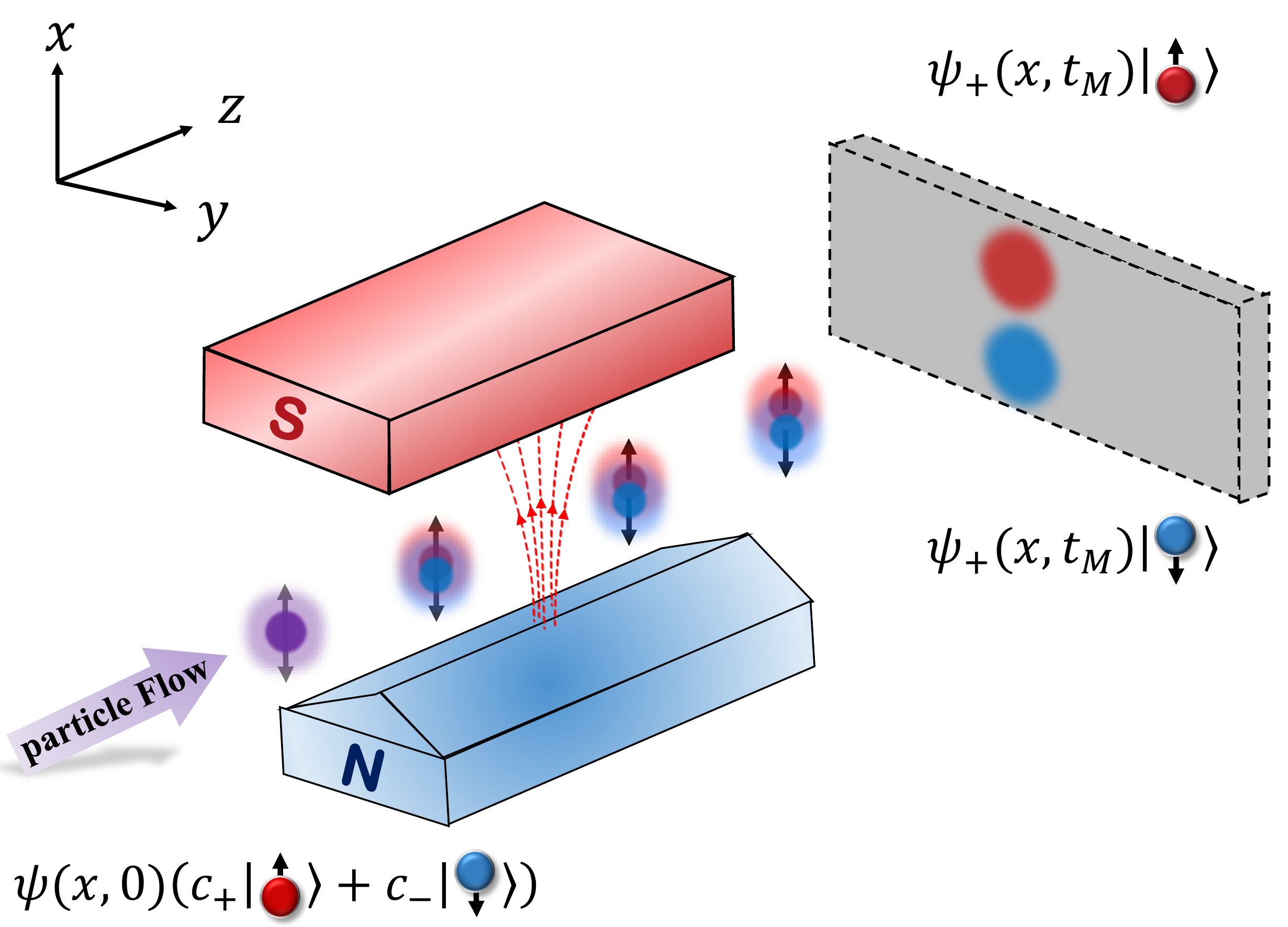}\caption{\label{fig:Schematic-diagram-of-1}Schematic illustration of Stern-Gerlach
experiment. The inhomogeneous magnetic field along $x$ direction
results in the entanglement between the spin degrees of freedom and
the spatial degrees of freedom of the particles, and thus separate
particles with different spins in space.}
\end{figure}

The Schematic illustration of the Stern-Gerlach experiment \citep{gerlach1922experimentelle,bohm2012quantum,platt1992modern,griffithsIntroductionQuantumMechanics2018}
is shown in Fig. \ref{fig:Schematic-diagram-of-1}. Non-interacting
particles with mass $m$, momentum $p$, and magnetic moment $\vec{\mu}$
are injected along $z$ direction with velocity $v_{z}$. Physically,
the applied inhomogeneous magnetic field $B(x)$ along $x$ direction
causes entanglement between the spin and the spatial degrees of freedom
of the particles. Such entanglement leads to quantum correlation between
these two kinds of degrees of freedom. 

In the ideal case where the spatial wave packets corresponding to
the spin-up and spin-down states of the particle are completely separated,
once the state of the spin (spatial) degrees of freedom is determined,
the state of the spatial (spin) degrees of freedom can be immediately
inferred. Nevertheless, in realistic cases, when the position of the
screen for observing the particle distribution is fixed, there is
usually overlap between the spatial wave packets associated with different
spin states. Intuitively, increasing the movement time of the particle
in the magnetic field (or the gradient of the magnetic field) will
make the spatial wave packets farther apart along the gradient of
the magnetic field, which results in a better distinction between
different spin degrees of freedom. In what follows, to quantify the
correlation between spin and spatial degrees of freedom, we first
study the evolution of the particle's spatial wave packet.

\subsection{\label{subsec:Time-evolution}Time evolution of the spatial wave
packet}

The Hamiltonian of the particle reads
\begin{equation}
\hat{H}=\dfrac{\hat{p}^{2}}{2m}-\mu_{x}B(x)\hat{\sigma}_{x},\label{eq:H}
\end{equation}
where $\hat{\sigma}_{x}=\left(|\uparrow\rangle\left\langle \uparrow\right|-|\downarrow\rangle\left\langle \downarrow\right|\right)$
is Pauli matrix in $x$ direction with the corresponding spin up (down)
state $|\uparrow\rangle(|\downarrow\rangle)$, and $\mu_{x}$ is the
$x$ component of the magnetic moment. Suppose the initial state of
particle in spin space and real space are non-entangled, and are respectively
$|s\rangle=c_{+}|\uparrow\rangle+c_{-}|\downarrow\rangle$ and $|\psi(x,0)\rangle$,
the full initial state of the particle is thus

\begin{equation}
|\Psi(0)\rangle=(c_{+}|\uparrow\rangle+c_{-}|\downarrow\rangle)\otimes|\psi(x,0)\rangle.
\end{equation}
In the coordinate representation, one has

\begin{equation}
\langle x|\Psi(0)\rangle=(c_{+}|\uparrow\rangle+c_{-}|\downarrow\rangle)\langle x|\psi(x,0)\rangle.
\end{equation}
According to the Schrodinger's equation, the time evolution of total
state of the particle follows as

\begin{align}
\langle x|\Psi(t)\rangle & =\langle x|e^{-i\hat{H}t/\hbar}|\Psi(0)\rangle\nonumber \\
 & =c_{+}\psi_{+}(x,t)|\uparrow\rangle+c_{-}\psi_{-}(x,t)|\downarrow\rangle,\label{eq:spatial}
\end{align}
where $\psi_{+}(x,t)$ and $\psi_{-}(x,t)$ are the normalized spatial
wave function corresponding to spin state $|\uparrow\rangle$ and
$|\downarrow\rangle$, respectively. Without loss of generality, we
focus on the the symmetric case that the superposition coefficients
of the initial states are $c_{+}=c_{-}=1/\sqrt{2}$.

In addition, we assume that the inhomogeneous magnetic field is slowly
varying over the range of particle motion, and hence $B(x)$ can be
expanded to the first-order of $x$ as $B(x)\approx B(0)+[(\partial B/\partial x)|_{x=0}]x$
\citep{platt1992modern}. In this sense, the Hamiltonian in Eq. (\ref{eq:H})
is approximated as

\begin{equation}
\hat{H}=\dfrac{\hat{p}^{2}}{2m}-fx\hat{\sigma}_{x},\label{eq:H-1}
\end{equation}
where $f\equiv\mu_{x}(\partial B/\partial x)|_{x=0}$ is the magnetic
force applied on the particle in classical picture. The spatial wave
function can be expressed with the path integral approach as \citep{feynmanQuantumMechanicsPath2010}

\begin{equation}
\psi_{\pm}(x,t)=\int_{-\infty}^{\infty}K_{\pm}(x,t;x',t')\psi(x',t'){\rm d}x',\label{eq:pathintegral}
\end{equation}
where $\psi(x',t')$ is the wave function at space-time coordinate
$(x',t')$, and $K_{\pm}(x,x';t)$ is the propagator determined by
the Hamiltonian of the system. In our case, without loss of generality,
we set initial time $t'=0$, and choose the initial wave packet as
a Gaussian wave packet with width $a$ as

\begin{equation}
\psi(x',0)\equiv\langle x|\psi(x',0)\rangle=(\dfrac{1}{2\pi a^{2}})^{1/4}e^{-\frac{x'^{2}}{4a^{2}}}.\label{eq:waveinitial}
\end{equation}
which is plotted in Fig. \ref{fig:Evolution-of-the} as the purple
solid curve in the $z=0$ plane. Note that the Hamiltonian in Eq.
(\ref{eq:H-1}) can be reduced to the coordinate space as

\begin{equation}
\hat{H}_{+}^{s}\equiv\left\langle \uparrow\right|\hat{H}|\uparrow\rangle=\dfrac{\hat{p}^{2}}{2m}-f\hat{x},\;\hat{H}_{-}^{s}\equiv\left\langle \downarrow\right|\hat{H}|\downarrow\rangle=\dfrac{\hat{p}^{2}}{2m}+f\hat{x},
\end{equation}
the propagator $K_{\pm}(x,x';t)\equiv\left\langle x\right|\exp(-i\hbar^{-1}\int_{0}^{t}\hat{H}_{\pm}^{s}dt)|x'\rangle$
associated with this linear potential reads \citep{feynmanQuantumMechanicsPath2010,hsuSternGerlachDynamicsQuantum2011}

\begin{equation}
K_{\pm}(x,x';t)=\sqrt{\frac{m}{2\pi i\hbar t}}{\rm exp}\left(-\frac{m(x-x')^{2}}{2i\hbar t}\pm\frac{f(x+x')t}{2i\hbar}+\frac{f^{2}t^{3}}{24i\hbar m}\right).\label{eq:K}
\end{equation}

Substituting Eqs. (\ref{eq:K}) and (\ref{eq:waveinitial}) into Eq.
(\ref{eq:pathintegral}), the wave function is obtained as

\begin{align}
\psi_{\pm}(x,t) & =(\dfrac{1}{2\pi a^{2}})^{1/4}\sqrt{\frac{m}{2\pi i\hbar t}}\int_{-\infty}^{\infty}{\rm exp}\left(-\frac{m(x-x')^{2}}{2i\hbar t}\pm\frac{f(x+x')t}{2i\hbar}+\frac{f^{2}t^{3}}{24i\hbar m}\right)e^{-\frac{x'^{2}}{4a^{2}}}{\rm d}x',\nonumber \\
 & =(\dfrac{1}{2\pi a^{2}})^{1/4}\sqrt{\frac{m}{2\pi i\hbar t}}\int_{-\infty}^{\infty}{\rm exp}\left(-\frac{x'^{2}}{4a^{2}}-\frac{m(x-x')^{2}}{2i\hbar t}\pm\frac{f(x+x')t}{2i\hbar}+\frac{f^{2}t^{3}}{24i\hbar m}\right){\rm d}x'.
\end{align}
Using the one-dimensional Gaussian integral 
\begin{equation}
\int_{-\infty}^{\infty}e^{-bz^{2}+cz}{\rm d}z=\sqrt{\frac{\pi}{b}}{\rm exp}\left(\frac{c^{2}}{4b}\right),{\rm Re({\it b})>0},\label{eq:Gaussian}
\end{equation}
the spatial wave function is straightforward calculated as

\begin{equation}
\psi_{\pm}(x,t)=\dfrac{(a^{2}/2\pi)^{1/4}}{\sqrt{a^{2}+i\hbar t/2m}}{\rm exp}\left\{ -\dfrac{if^{2}t^{3}}{6\hbar m}-\dfrac{[x\mp ft^{2}/(2m)]^{2}}{4(a^{2}+i\hbar t/2m)}\pm\frac{iftx}{\hbar}\right\} ,\label{eq:psi_pm}
\end{equation}

\begin{flushleft}
The norms of the above wave functions at $t=0,3,6$ are illustrated
in Fig. \ref{fig:Evolution-of-the}, where (a) is plotted with $f=0.5$
and (b) is plotted with $f=1.5$, respectively. All other parameters
are set to $1$ in this example. The red dash-dotted curves and blue
dashed curves are associated with $\psi_{+}(x,t)$ and $\psi_{-}(x,t)$,
respectively. It is clearly seen in this figure that the degree of
separation of the wave packets corresponding to different spin states
increases with time. And for a given $t$, stronger magnetic field
gradient, i.e., $f\uparrow$, causes the wave packets to be separated
farther apart.
\par\end{flushleft}

\begin{center}
\begin{figure}
\begin{centering}
\subfloat[]{\includegraphics[width=8.5cm]{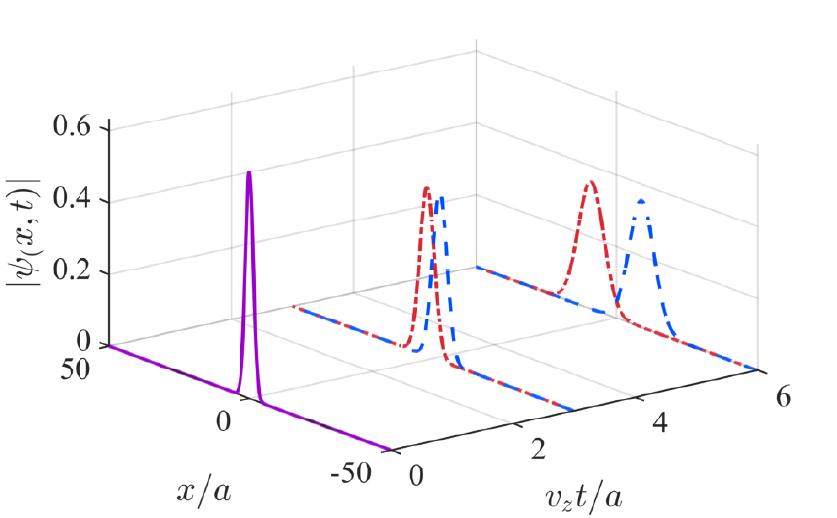}

}\subfloat[]{\includegraphics[width=8.5cm]{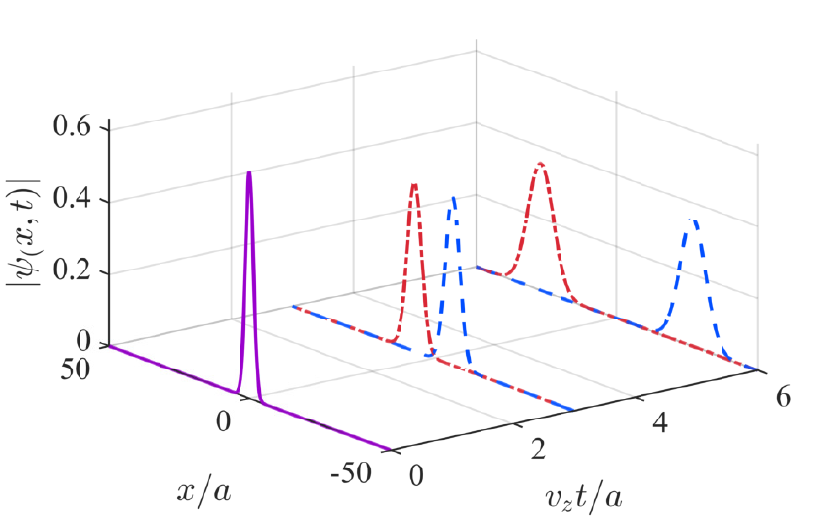}

}
\par\end{centering}
\caption{\label{fig:Evolution-of-the}Evolution of the spatial wave packet
with $f=0.5$(a) and $f=1.5$(b). The purple solid curve represents
the initial wave packet in Eq. (\ref{eq:waveinitial}). The red dash-dotted
(blue dashed) curves mark $|\psi_{+}(x,t)|$ ($|\psi_{-}(x,t)|$)
at $t=3$ and $t=6$. In this figure, we choose $a=m=\hbar=1$, the
velocity of the particle in the $z$ axis is set to $v_{z}=1$.}
\end{figure}
\par\end{center}

After the measurement process of duration $t_{{\rm M}}$, the spacial
overlapping integral $F\equiv|\langle\psi_{+}|\psi_{-}\rangle|$ is 

\begin{align}
F(t_{{\rm M}}) & =\int_{-\infty}^{\infty}\psi_{+}^{*}(x,t_{{\rm M}})\psi_{-}(x,t_{{\rm M}})dx\nonumber \\
 & =\frac{\left(a^{2}/2\pi\right)^{1/2}}{\sqrt{a^{4}+\hbar^{2}t_{M}^{2}/4m^{2}}}\int_{-\infty}^{\infty}{\rm exp}\left\{ -\dfrac{[x-ft_{M}^{2}/(2m)]^{2}}{4(a^{2}-i\hbar t_{{\rm M}}/2m)}-\dfrac{[x+ft_{M}^{2}/(2m)]^{2}}{4(a^{2}+i\hbar t_{{\rm M}}/2m)}-2\frac{ift_{{\rm M}}x}{\hbar}\right\} dx\nonumber \\
 & =\frac{\left(a^{2}/2\pi\right)^{1/2}}{\sqrt{a^{4}+\hbar^{2}t_{M}^{2}/4m^{2}}}\int_{-\infty}^{\infty}{\rm exp}\left\{ -\dfrac{a^{2}\left(x^{2}+f^{2}t_{M}^{4}/4m^{2}\right)-i\hbar fxt_{M}^{3}/2m^{2}}{2(a^{4}+\hbar^{2}t_{M}^{2}/4m^{2})}-2\frac{ift_{{\rm M}}x}{\hbar}\right\} dx,
\end{align}
which is straightforward calculated as, by using Eq. (\ref{eq:Gaussian}),

\begin{equation}
F(\tilde{t})={\rm exp}\left[-\left(2\alpha^{2}\tilde{t}^{2}+\frac{\alpha^{2}\tilde{t}^{4}}{8}\right)\right].\label{eq:Ft}
\end{equation}
Here, $\tilde{t}\equiv t_{{\rm M}}/\tau_{\mathrm{M}}$ is the dimensionless
measurement time, $\tau_{\mathrm{M}}\equiv ma^{2}/\hbar$ is the characteristic
time of the measurement process. $\text{\ensuremath{\alpha}}\equiv ma^{3}f/\hbar^{2}$,
which characterizes the strength of decoherence after measurement,
is proportional to the magnetic force $f$. For given $\tilde{t}$,
a stronger magnetic field (larger $f$) results in a smaller $F(t_{{\rm M}})$,
i.e. a more pronounced decoherence corresponds to farther apart wave
packets in Fig. \ref{fig:Evolution-of-the}. Obviously, $F(\tilde{t})\rightarrow0$
in the long-time limit $\tilde{t}\to\infty$ of the measurement process.
In this case, there is no overlapping between $\psi_{+}(x,t)$ and
$\psi_{-}(x,t)$, and the spin of the particle can be perfectly distinguished
by observing its spatial wave function in $x$ direction. Therefore,
an ideal measurement of the spin is achieved in the Stern-Gerlach
experiment.

\subsection{\label{subsec:Conditional-spatial-distribution-probability}Conditional
spatial distribution probability}

To discuss the non-ideal measurement within finite measuring duration,
we first study the conditional probability $p(L|\uparrow)$ that a
spin-up particle appears in the ``left'' area of the screen, and
the conditional probability $p(R|\downarrow)$ that a spin-down particle
appears in the right'' area of the screen. Here, left and right are
divided with respect to the $x$-direction, which corresponds to the
area with $x>0$ and $x<0$, respectively. By definitions, the probabilities
are

\begin{align}
p(L|\uparrow)\equiv & \int_{0}^{\infty}\langle x|\left[\langle\uparrow|\Psi(t)\rangle\langle\Psi(t)|\uparrow\rangle\right]|x\rangle dx\nonumber \\
= & \int_{0}^{\infty}\left|\psi_{+}(x,t_{{\rm M}})\right|^{2}\text{{\rm d}}x,\label{eq:p(R_up)}
\end{align}
and

\begin{align}
p(R|\downarrow)\equiv & \int_{-\infty}^{0}\langle x|\left[\langle\downarrow|\Psi(t)\rangle\langle\Psi(t)|\downarrow\rangle\right]|x\rangle dx\nonumber \\
= & \int_{-\infty}^{0}\left|\psi_{-}(x,t_{{\rm M}})\right|^{2}\text{{\rm d}}x.
\end{align}
Obviously, due to the symmetry of the spatial wave functions obtained
in Eq. (\ref{eq:psi_pm}), the above two probabilities equals to each
other, i.e., $p(L|\uparrow)=p(R|\downarrow)\equiv p(t_{{\rm M}})$.
Substituting Eq. (\ref{eq:psi_pm}) into Eq. (\ref{eq:p(R_up)}) yields 

\begin{align}
p\left(t_{{\rm M}}\right) & =\int_{0}^{\infty}\left|\psi_{+}(x,t_{{\rm M}})\right|^{2}\text{{\rm d}}x\nonumber \\
 & =\frac{\left(a^{2}/2\pi\right)^{1/2}}{\sqrt{a^{4}+\hbar^{2}t_{{\rm M}}^{2}/4m^{2}}}\int_{0}^{\infty}{\rm exp}\left[-\frac{a^{2}}{2\left(a^{4}+\hbar^{2}t_{{\rm M}}^{2}/4m^{2}\right)}\left(x-\frac{ft_{{\rm M}}^{2}}{2m}\right)^{2}\right]\text{{\rm d}}x\nonumber \\
 & =\frac{\left(a^{2}/2\pi\right)^{1/2}}{\sqrt{a^{4}+\hbar^{2}t_{{\rm M}}^{2}/4m^{2}}}\int_{-\frac{ft_{{\rm M}}^{2}}{2m}}^{\infty}{\rm exp}\left[-\frac{a^{2}}{2\left(a^{4}+\hbar^{2}t_{{\rm M}}^{2}/4m^{2}\right)}y^{2}\right]\text{{\rm d}}y\label{eq:Ptappendix}
\end{align}
Using the Gauss error function 
\begin{equation}
{\rm erf}\left(z\right)\equiv\frac{2}{\sqrt{\pi}}\int_{0}^{z}e^{-u^{2}}{\rm d}u,
\end{equation}
we simplify Eq. (\ref{eq:Ptappendix}) as

\begin{equation}
p\left(t_{{\rm M}}\right)=p(\tilde{t})=\frac{1}{2}\left[1+{\rm erf}\left(\frac{\alpha\tilde{t}^{2}}{\sqrt{2\tilde{t}^{2}+8}}\right)\right],\label{eq:P(T)}
\end{equation}
where ${\rm erf}(z)=2\pi^{-1/2}\int_{0}^{z}\mathrm{exp}(-u^{2})\text{{\rm d}}u$
is the Gauss error function. When the dimensionless measurement time
$\tilde{t}$ approaches infinity and zero we have $\lim_{\tilde{t}\rightarrow\infty}p(\tilde{t})=1$
and $\lim_{\tilde{t}\rightarrow0}p(\tilde{t})=1/2$ , respectively.
Correspondingly, the probability of a spin-up particle appears in
the ``left'' area, and the probability of a spin-down particle appears
in the \textquotedblleft right'' area are
\begin{equation}
p(R|\uparrow)=\int_{-\infty}^{0}\left|\psi_{+}(x,t_{{\rm M}})\right|^{2}\text{{\rm d}}x=1-p(\tilde{t}),\label{eq:Prup}
\end{equation}
and

\begin{equation}
p(L|\downarrow)=\int_{0}^{\infty}\left|\psi_{-}(x,t_{{\rm M}})\right|^{2}\text{{\rm d}}x=1-p(\tilde{t}).\label{eq:Pldown}
\end{equation}
Here the normalized condition for conditional probabilities $p(L|\theta)+p(R|\theta)=1,\theta\in\{\uparrow,\downarrow\}$
has been used.

In the long-time regime of $\tilde{t}\gg1$, we have $p(\tilde{t})\rightarrow1$,
which results in the conditional probabilities $p(L|\uparrow)=p(R|\downarrow)\rightarrow1.$
In this case, we can achieve an ideal measurement for the spin state
through the measurement of the space state: when a particle is observed
to be on the left (right) side of the screen, we can definitely determine
that its spin state is $|\uparrow\rangle$ ($|\downarrow\rangle$).
Conversely, we can also determine the spatial position of a particle
by measuring its spin state. However, for an arbitrary $\tilde{t}$
away from the long-time regime, even if the spin state of the particle
has been accurately obtained, we can only infer the spatial state
of the particle probabilistically. For example, when we find the spin
state of the particle is $|\uparrow\rangle$, we therefore infer that
the particle is on the left side of the screen. According to Eq. (\ref{eq:Prup}),
the probability that this inference is correct is $p_{c}=p(\tilde{t})$.
On the contrary, the probability of this inference being wrong is
$p_{w}=1-p(t)$, which corresponds to the case where the particle
is actually on the right side of the screen.

\subsection{\label{subsec:Correlation-between-spin-state-and-spatial-state}Measurement
ideality}

As we mentioned in the main text, the measurement of the position
of the particle by MD is realized through the detection of the spin
state. After the measurement, we assume that the instrument detects
the spin perfectly, then the MD state $|D_{\theta}\rangle$ becomes
the relative state of the spin state $|\theta\rangle$, and the conditional
probability of spin state given MD state $p(\theta|D)$ satisfies
\begin{equation}
p(\uparrow|D_{\uparrow})=p(\downarrow|D_{\downarrow})=1,p(\uparrow|D_{\downarrow})=p(\downarrow|D_{\uparrow})=0
\end{equation}

If the observation time is not long enough (in comparison with $\tau_{M}$),
the quantum measurement of the position of the particle is non-ideal.
We quantify the correlation information of the demon and the particle
with the mutual entropy \citep{kullback1997information,2002Quantum,li2017production,Still2020PRL.124.050601,ma2021works}

\begin{equation}
I=S_{P}+S_{D}-S_{DP}.
\end{equation}
Here, $S_{Y}=-\sum_{y\in Y}p(y){\rm ln}p(y)$ is the Shannon entropy
with respect to the variable $Y$, $S_{DP}=-\sum_{P}\sum_{D}p\left(P,D\right){\rm ln}p\left(P,D\right)$
is the joint entropy of the particle and the demon, where the joint
probability is $p\left(P,D\right)=p(P|D)p(D)$. Using such definition
and Eqs. (\ref{eq:Prup}) and (\ref{eq:Pldown}), the mutual information
is obtained as

\begin{align}
I & =S_{P}+S_{D}-S_{DP}\nonumber \\
 & =\sum_{P\in\{R,L\}}\sum_{D\in\{D_{\uparrow},D_{\downarrow}\}}p(P,D){\rm log}\frac{p(P|D)}{p(P)}\nonumber \\
 & =\sum_{P\in\{R,L\}}\sum_{D\in\{D_{\uparrow},D_{\downarrow}\}}\sum_{\theta\in\{\uparrow,\downarrow\}}p(P|D)p(D){\rm log}\frac{p(P|D)}{p(P)}\\
 & =\sum_{P\in\{R,L\}}\sum_{D\in\{D_{\uparrow},D_{\downarrow}\}}\sum_{\theta\in\{\uparrow,\downarrow\}}p(P|\theta)p(\theta|D)p(D){\rm log}\frac{p(P|\theta)p(\theta|D)}{p(P)}\\
 & =p(R|\uparrow)p(\uparrow){\rm log}\frac{p(R|\uparrow)}{p(R)}+p(R|\downarrow)p(\downarrow){\rm log}\frac{p(R|\downarrow)}{p(R)}+p(L|\uparrow)p(\uparrow){\rm log}\frac{p(L|\uparrow)}{p(L)}+p(L|\downarrow)p(\downarrow){\rm log}\frac{p(L|\downarrow)}{p(L)}\nonumber \\
 & ={\rm ln}2+p(\tilde{t}){\rm ln}p(\tilde{t})+\left[1-p(\tilde{t})\right]{\rm ln}\left[1-p(\tilde{t})\right],\label{eq:I_PD}
\end{align}
where $p(\uparrow)=p(\downarrow)=1/2$ has beenused according to the
symmetric choose of superposition coefficients in the initial states. 

According to the above equation and Eq. (\ref{eq:P(T)}), for the
example shown in Fig. \ref{fig:Evolution-of-the} (a) with $f=0.5$,
$I_{\tilde{t}=3}=0.355$ and $I_{\tilde{t}=6}=0.677$; while for Fig.
\ref{fig:Evolution-of-the} (b) with $f=1.5$, $I_{\tilde{t}=3}=0.692$
and $I_{\tilde{t}=6}=0.693$. Moreover, we define the measurement
ideality as $\text{\ensuremath{\mathcal{M}(\tilde{t})\equiv I(\tilde{t})/{\rm ln}2}}$,
namely,

\begin{equation}
\mathcal{M}(\tilde{t})=1+\frac{p(\tilde{t}){\rm ln}p(\tilde{t})+\left[1-p(\tilde{t})\right]{\rm ln}\left[1-p(\tilde{t})\right]}{{\rm ln}2}
\end{equation}
which is illustrated in Fig.2 of the main text.

The long-time and short-time behaviors of the mutual entropy are discussed
below. In the long-time regime of $\tilde{t}\gg1$, we have $p\left(\tilde{t}\right)\rightarrow1$,
keeping to the first order of $1-p\left(\tilde{t}\right)$, we have

\begin{align}
\mathcal{M}\left(\tilde{t}\gg1\right) & \approx1+\frac{{\rm ln}\left[1+p\left(\tilde{t}\right)-1\right]}{{\rm ln}2}\nonumber \\
 & \approx1+\frac{p\left(\tilde{t}\right)-1}{{\rm ln}2}\nonumber \\
 & =1+\frac{1}{2{\rm ln}2}\left[{\rm erf}\left(\frac{\alpha\tilde{t}^{2}}{\sqrt{8+\tilde{t}^{2}}}\right)-1\right]\nonumber \\
 & \approx1-\frac{1-{\rm erf}\left(\alpha\tilde{t}\right)}{2{\rm ln}2}.
\end{align}
On the other hand, in the short-time regime of $\tilde{t}\ll1$, we
have

\begin{align}
p\left(\tilde{t}\ll1\right) & =\frac{1}{2}\left[{\rm 1+erf}\left(\frac{\alpha\tilde{t}^{2}}{\sqrt{8+\tilde{t}^{2}}}\right)\right]=\frac{1}{2}+\frac{1}{\sqrt{\pi}}\int_{0}^{\frac{\alpha\tilde{t}^{2}}{\sqrt{8+\tilde{t}^{2}}}}e^{-u^{2}}{\rm d}u\nonumber \\
 & \approx\frac{1}{2}+\frac{1}{\sqrt{\pi}}\int_{0}^{\frac{\alpha\tilde{t}^{2}}{2\sqrt{2}}}(1-u^{2}){\rm d}u\nonumber \\
 & \approx\frac{1}{2}+\frac{\alpha\tilde{t}^{2}}{2\sqrt{2\pi}}.
\end{align}
Then the mutual entropy is approximated as

\begin{align}
\mathcal{M}\left(\tilde{t}\ll1\right) & =2+\left(\frac{1}{2{\rm ln}2}+\frac{\alpha\tilde{t}^{2}}{2{\rm ln}2\sqrt{2\pi}}\right){\rm ln}\frac{\frac{1}{2}+\frac{\alpha\tilde{t}^{2}}{2\sqrt{2\pi}}}{2}+\left(\frac{1}{2{\rm ln}2}-\frac{\alpha\tilde{t}^{2}}{2{\rm ln}2\sqrt{2\pi}}\right){\rm ln}\left[\frac{\frac{1}{2}-\frac{\alpha\tilde{t}^{2}}{2\sqrt{2\pi}}}{2}\right]\nonumber \\
 & =\left(\frac{1}{2{\rm ln}2}+\frac{\alpha\tilde{t}^{2}}{2{\rm ln}2\sqrt{2\pi}}\right){\rm ln}\left(1+\frac{\alpha\tilde{t}^{2}}{\sqrt{2\pi}}\right)+\left(\frac{1}{2{\rm ln}2}-\frac{\alpha\tilde{t}^{2}}{2{\rm ln}2\sqrt{2\pi}}\right){\rm ln}\left(1-\frac{\alpha\tilde{t}^{2}}{\sqrt{2\pi}}\right)\nonumber \\
 & \approx\left(\frac{1}{2{\rm ln}2}+\frac{\alpha\tilde{t}^{2}}{2{\rm ln}2\sqrt{2\pi}}\right)\left[\frac{\alpha\tilde{t}^{2}}{\sqrt{2\pi}}-\frac{1}{2}\left(\frac{\alpha\tilde{t}^{2}}{\sqrt{2\pi}}\right)^{2}\right]+\left(\frac{1}{2{\rm ln}2}-\frac{\alpha\tilde{t}^{2}}{2{\rm ln}2\sqrt{2\pi}}\right)\left[-\frac{\alpha\tilde{t}^{2}}{\sqrt{2\pi}}-\frac{1}{2}\left(\frac{\alpha\tilde{t}^{2}}{\sqrt{2\pi}}\right)^{2}\right]\nonumber \\
 & =\frac{\alpha^{2}\tilde{t}^{4}}{8\pi{\rm ln}2}=\frac{m^{2}a^{6}f^{2}\tilde{t}^{4}}{8\pi{\rm ln}2\hbar^{4}},
\end{align}
which is proportional to $\tilde{t}^{4}$. The proportional coefficient
increases with the square of the magnetic force and with the sixth
power of the initial width of the wave packet.

\section{\label{sec:Thermodynamic-signature-of-full-finite-time-description}Thermodynamic
signature of full finite-time description of quantum Szilard heat
engine}

In the main text, we mainly focus on the finite-time effect in the
measurement process (Stage I). In this section, we discuss the finite-time
behaviors of the other processes (Stages II and III) in the engine
cycle.

\subsection{\label{subsec:Information-recording-and-Information-erasing}Thermodynamic
signatures of information recording and information erasing process}

In this part, we further consider the finite-time effect in both the
information recording and information erasing processes. According
to the finite-time Landauer's principle \citep{berut2012experimental,proesmansFiniteTimeLandauerPrinciple2020,ma2022minimal},
the non-ideal erasing process leads to irreversible energy dissipation
which is inversely proportional to the erasing time $t_{{\rm E}}$,
namely,

\begin{equation}
W_{\mathrm{E}}(t_{\mathrm{E}})=k_{\mathrm{B}}T_{{\rm C}}{\rm ln}2\left(1+\frac{C_{{\rm E}}}{t_{\mathrm{E}}}\right),\label{eq:WE-1}
\end{equation}
where $C_{{\rm E}}$ is a constant determined by specific erasing
process \citep{proesmansFiniteTimeLandauerPrinciple2020,maMinimalEnergyCost2022}.
In this manner, we rewrite the total work of the QSE performs in one
cycle as
\begin{equation}
W=\mathcal{M}(t_{{\rm M}})k_{{\rm B}}T_{{\rm H}}{\rm ln}2-k_{\mathrm{B}}T_{{\rm C}}{\rm ln}2\left(1+\frac{C_{{\rm E}}}{t_{\mathrm{E}}}\right)\label{eq:W-1}
\end{equation}
Then, the power and efficiency of the information heat engine becomes

\begin{equation}
\eta(t_{{\rm M}},t_{{\rm E}})=\frac{W}{\mathcal{M}(t_{{\rm M}})k_{{\rm B}}T_{{\rm H}}{\rm ln}2}=1-\frac{1-\eta_{{\rm C}}}{\mathcal{M}(t_{{\rm M}}){\rm ln}2}\left({\rm ln}2+\frac{C_{{\rm E}}}{t_{\mathrm{E}}}\right),\label{eq:eta-1}
\end{equation}
and

\begin{equation}
P(t_{{\rm M}},t_{{\rm E}})=\frac{\mathcal{M}(t_{{\rm M}})k_{{\rm B}}T_{{\rm H}}{\rm ln}2-k_{\mathrm{B}}T_{{\rm C}}{\rm ln}2\left(1+C_{{\rm E}}t_{\mathrm{E}}^{-1}\right)}{t_{{\rm M}}+t_{{\rm E}}}\label{eq:P}
\end{equation}

\begin{figure}
\centering{}\includegraphics[width=8.5cm]{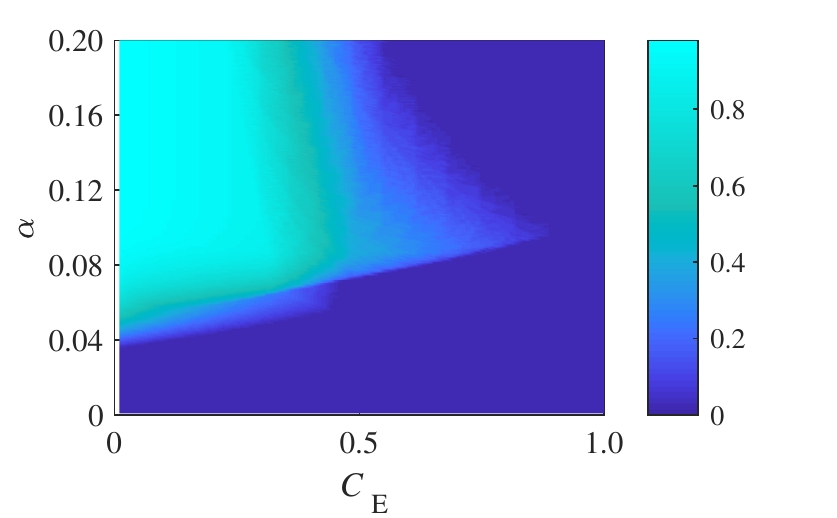} \caption{\label{fig:EMP-1} The width of the range of $\eta_{{\rm C}}$ that
the efficiency at maximum power(EMP) can exceed $\eta_{+}$ in two-dimensional
parameter space $\left(\alpha,C_{{\rm E}}\right)$. Lighter area corresponds
to wider range of $\eta_{{\rm C}}$ within which EMP can exceed the
typical bound of EMP $\eta_{+}=\eta_{\mathrm{C}}/(2-\eta_{\mathrm{C}})$
.}
\end{figure}

Similar to the situation where only the finite-time effect of the
information recording process is taken into account, the EMP of the
engine in this situation can also exceed the upper limit of the efficiency
of low-dissipation Carnot heat engine $\eta_{+}=\eta_{\mathrm{C}}/(2-\eta_{\mathrm{C}})$
at certain range of $\eta_{{\rm C}}$. In order to quantify the influence
of parameters $\alpha$ and $C_{{\rm E}}$ on the EMP, in Fig. \ref{fig:EMP-1},
we show the width of the range of $\eta_{C}$ within which EMP can
exceed $\eta_{+}$ in the two-dimensional parameter space $\left(\alpha,C_{{\rm E}}\right)$.
It can be seen from the figure that wider range of $\eta_{{\rm C}}$
requires relative large value of $\alpha$ and small value of $C_{{\rm E}}$,
which consistent well with physical intuition: for a given operation
time, a larger $\alpha$ results in a more complete recording of information,
while a smaller $C_{{\rm E}}$ corresponds to a more reversible erasing
process with less energy dissipation.

\begin{figure}
\centering{}\includegraphics[width=8.5cm]{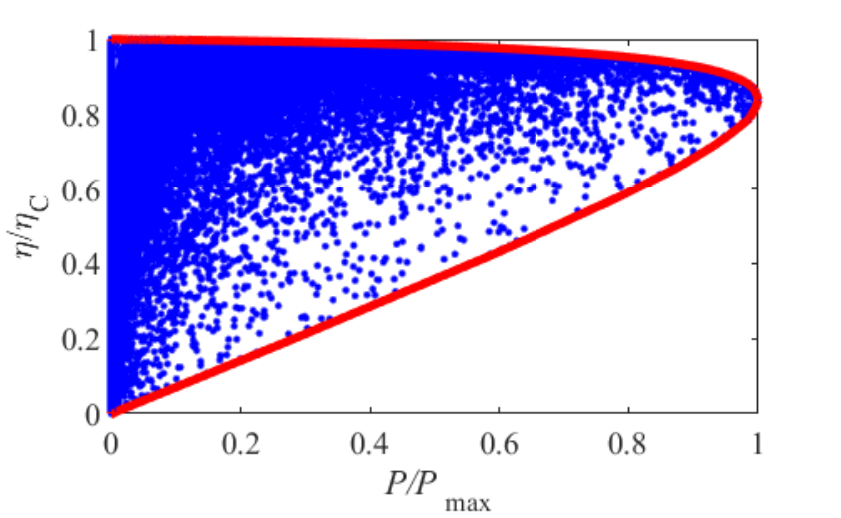}\caption{\label{fig:E-P tradeoff2} Power-efficiency trade-off relation of
the information heat engine, where the finite-time effects of both
the information recording and information erasing process are considered.
The Blue dots are obtained numerically from Eqs. (\ref{eq:eta-1})
and (\ref{eq:P}) with $10^{7}$ sets of measurement time and erasing
time pair $\left(\tilde{t},t_{{\rm E}}\right)$ chosen randomly from
$\tilde{t}\in\left(1,4\right)$ and $t_{{\rm E}}\in\left(1,400\right)$.
The Red solid curve marks the envelop of these blue dots, namely,
the boundary of the power-efficiency trade-off. In this plot, we choose
$\eta_{{\rm C}}=0.6$, $\alpha=0.4$ and $C_{{\rm E}}=0.5$.}
\end{figure}

The power-efficiency trade-off relation is shown in Fig. \ref{fig:E-P tradeoff2}
by calculating the efficiency $\eta/\eta_{\mathrm{C}}$ and the power
$P/P_{\mathrm{max}}$ corresponding to $10^{7}$ of random measurement
time and erasing time pairs $\left(\tilde{t},t_{{\rm E}}\right)$
from$\left(1,1\right)$ to $\left(4,400\right)$. Each blue dot in
Fig. \ref{fig:E-P tradeoff2} represents a heat engine cycle with
different $\left(\tilde{t},t_{{\rm E}}\right)$, and the red solid
curve is the envelop of these blue dots which shows the boundary of
the trade-off relation. Different from the single-variable case shown
in Fig. 5(b) in the main text where the system can only operate with
power and efficiency locate on the curve, the heat engine with the
finite-time effects in both the information recording and information
erasing processes taken into consideration can operate at anywhere
within the envelop. The efficiency at an arbitrary given power $P_{a}$
is bounded by two intersection points of the vertical line $P=P_{a}$
and the red solid curve. As $P/P_{\mathrm{max}}\to0$, $\eta/\eta_{{\rm C}}$
approaches $1$, which covers the quasi-static regime of the information
heat engine.

\subsection{\label{subsec:Work-ouputting-and-Information-erasing}Thermodynamic
signature of work outputting and information erasing process}

In the last case to be studied, we assume $\tau_{\mathrm{M}}\ll t_{{\rm M}}\ll t_{{\rm O}}(t_{{\rm E}})$.
In this regime, the information recording is complete with $\mathcal{M}(t_{{\rm M}})\approx1$,
and operation time of the measurement process can be ignored in comparison
with that of the work outputting and information erasing process.
Below, we focus on the EMP of the information engine.

In the work outputting process, the information engine is contact
with a heat reservoir of temperature $T_{{\rm H}}$, and such a process
is the same as the finite-time isothermal process in the Carnot-like
cycle \citep{CA,schmiedl2008efficiency,EspositoPRL2010,Constraintrelationyhma}.
In the low-dissipation regime \citep{EspositoPRL2010,Constraintrelationyhma,yhmaoptimalcontrol,2020Optimal,ma2020experimental},
the output work follows as

\begin{equation}
W_{\mathrm{O}}\left(t_{\mathrm{O}}\right)=T_{{\rm H}}\Delta S\left(1-\frac{C_{\mathrm{O}}}{t_{\mathrm{O}}}\right),\label{eq:WO}
\end{equation}
where $\Delta S=k_{\mathrm{B}}\ln2$ is the reversible entropy change
\citep{dong2011quantum,kim2011quantum} and $C_{\mathrm{O}}$ is the
dissipation constant. It follows from Eqs. (\ref{eq:WO}) and (\ref{eq:W-1})
that the power of engine becomes a function of $t_{\mathrm{O}}$ and
$t_{\mathrm{E}}$ as

\begin{align}
P & \left(t_{\mathrm{O}},t_{\mathrm{E}}\right)=\frac{W_{\mathrm{O}}\left(t_{\mathrm{O}}\right)-W_{\mathrm{E}}\left(t_{\mathrm{E}}\right)}{t_{\mathrm{O}}+t_{\mathrm{E}}}\\
 & =T_{{\rm H}}\Delta S\frac{\eta_{\mathrm{C}}-C_{{\rm O}}t_{\mathrm{O}}^{-1}-\left(1-\eta_{\mathrm{C}}\right)C_{{\rm E}}t_{\mathrm{E}}^{-1}}{t_{\mathrm{O}}+t_{\mathrm{E}}}.
\end{align}
It is easy to check that the maximum power $P=P_{\mathrm{max}}$ is
achieved when $\partial P/\partial t_{\mathrm{O}}=0$ and $\partial P/\partial t_{\mathrm{E}}=0$,
and the optimal operation times at maximum power are obtained as

\begin{equation}
t_{\mathrm{O}}^{*}=\frac{2C_{\mathrm{O}}}{\eta_{\mathrm{C}}}\left(1+\sqrt{\frac{C_{\mathrm{E}}}{C_{\mathrm{O}}}}\right),\;t_{\mathrm{E}}^{*}=t_{\mathrm{O}}^{*}\sqrt{\frac{C_{\mathrm{E}}}{C_{\mathrm{O}}}}
\end{equation}
Correspondingly, the efficiency at maximum power (EMP) for this information
engine is

\begin{equation}
\eta_{MP}=1-\frac{W_{\mathrm{E}}\left(t_{\mathrm{E}}^{*}\right)}{W_{\mathrm{O}}\left(t_{\mathrm{O}}^{*}\right)}=\frac{\eta_{\mathrm{C}}}{2-\kappa\eta_{\mathrm{C}}}.\label{eq:EMP}
\end{equation}
Here $\kappa=\left[1+\sqrt{C_{\mathrm{E}}/\left(C_{\mathrm{O}}\right)}\right]^{-1}$
depends on the asymmetry of dissipation when the heat engine is in
contact with two heat reservoirs. Since $0\leq\kappa\leq1$ for different
$\Sigma_{\mathrm{c}}/\Sigma_{\mathrm{h}}$, $\eta_{MP}$ is found
to satisfy the following inequality

\begin{equation}
\eta_{\mathrm{L}}\equiv\frac{\eta_{\mathrm{C}}}{2}\leq\eta_{MP}\leq\frac{\eta_{\mathrm{C}}}{2-\eta_{\mathrm{C}}}\equiv\eta_{\mathrm{U}}.\label{eq:inequality}
\end{equation}
In the above relation, $\eta_{\mathrm{U}}$ and $\eta_{\mathrm{L}}$
are respectively the upper and lower bound of EMP. 

Equation (\ref{eq:inequality}) covers the result obtained with the
low-dissipation model for finite-time Carnot engine \citep{EspositoPRL2010}.
This indicates that when the information recording is ideal and the
corresponding duration is ignored, the finite-time Szilard engine
is mapped to a low-dissipation Carnot heat engine. When the dissipation
coefficients in the work outputting process and information erasing
process are the same as each other, i.e., $C_{\mathrm{E}}/C_{\mathrm{O}}=1$,
the EMP of Eq. (\ref{eq:EMP}) recovers the so-called Curzon-Ahlborn
efficiency, i.e., $\eta_{\mathrm{CA}}=1-\sqrt{T_{\mathrm{C}}/T_{\mathrm{H}}}$
\citep{CA}.

\bibliographystyle{apsrev}
\bibliography{FTOSHE}